\documentstyle[11pt,newpasp,twoside,epsf]{article}
\def\edcomment#1{\iffalse\marginpar{\raggedright\sl#1\/}\else\relax\fi}
\reversemarginpar

\begin{document}
\title{Low Redshift Lyman $\alpha$ absorbers and their Connection with 
Galaxies}
\author{Simon L. Morris}
\affil{Physics Department, University of Durham, South Rd., Durham DH1 3LE, UK}
\author{Buell Jannuzi}
\affil{National Optical Astronomy Observatories, 950 North Cherry Avenue, P.O. 
Box 26732, Tucson, Arizona 85726}
\author{Ray Weymann}
\affil{Carnegie Observatories, 813 Santa Barbara St., Pasadena, CA 91101}

% Add as many "\author" and "\affil" fields as necessary.

\begin{abstract}
We review the ongoing debate about the relationship between low redshift Lyman 
$\alpha$ absorbers and luminous galaxies. In particular, we discuss the 
difficulty of `assigning' a particular absorber to a particular galaxy, and 
consider methods of circumventing this problem. We also provide a status report 
on an ongoing project collecting more data to address this issue, and show some 
results for a close together pair of QSOs providing two adjacent lines of sight 
through the inter-galactic medium.
\end{abstract}

\section{Introduction}
Because the analysis of the new data taken for this project is not complete, we 
will use the first part of this contribution to explain and (we hope) clarify 
the ongoing debate about the relationship between low redshift Lyman $\alpha$ 
absorbers and galaxies. In the second part of the contribution, we will give a 
status report on the new data analysis.

A simplified version of the two sides of the debate might be that either (a) 
all low redshift Lyman $\alpha$ absorbers are part of physically-distinct 
luminous-galaxy halos, or (b) that they are all part of the filamentary 
structure seen in recent SPH/Mesh structure formation models, and are only 
related to galaxies by the fact that they both are following the underlying 
dark matter distribution. Both of these positions have been vigorously defended 
in the literature, and at this conference. Proponents of (a) include: Lanzetta 
et al. (1995); Lanzetta, Webb \& Barcons (1996); Barcons et al. (1998); Chen et 
al. (1998); Ortiz-Gil et al. (1999); Chen, Lanzetta and Fern{\'a}ndez-Soto 
(2000); and Linder (2000). Proponents of (b) include: Morris et al. (1991); 
Morris et al. (1993); Mo \& Morris (1994); Morris \& van den Bergh (1994); 
Weymann et al. (1995); Dinshaw et al. (1995); Stocke et al. (1995); Rauch, 
Weymann \& Morris (1996); Shull, Stocke \& Penton (1996); van Gorkom et al. 
(1996); Bowen, Blades \& Pettini (1996); Dinshaw et al. (1997); Jannuzi et al. 
(1998); Impey, Petry \& Flint (1999); Penton, Shull \& Stocke (2000); and 
Penton, Stocke \& Shull (2000). Finally to complete the reference list, some of 
the SPH/mesh models which include mention of the Lyman $\alpha$ absorbers are: 
Hernquist et al. (1996); Cen et al. (1998); Dav{\'e} et al. (1999); Cen \& 
Ostriker (1999); and Dav{\'e} \& Tripp (2001).

Before embarking on a discussion of which hypothesis is correct, it should be 
emphasized that even if one believes that one can obtain a clear answer (either 
(a) or (b) above), that the answer must be a function of column density. At 
column densities N$_{H}>10^{21}$ cm$^{-2}$, one is probing a thickness of 
material comparable to our own galaxy's disk, and hence something almost 
certainly part of a galaxy, at least at low redshift. At column densities 
N$_{H}\sim10^{12}$ cm$^{-2}$, one is getting close to the expected neutral 
hydrogen content of fluctuations within voids, and must expect a much more 
heterogeneous set of causes for the absorption.

\section{Philosophy}
In order to frame the above debate, it is useful first to list what the 
observables are. In some sense, any question that cannot be addressed using 
these observables is uninteresting.

\noindent
Measurable Absorber Properties:
 
\begin{itemize}
 \item redshift
 \item x,y on sky
 \item absorption line properties such as Equivalent Width (EW), doppler b 
parameter, column density, and deviations from a Voigt profile
\end{itemize}

\noindent
Measurable Galaxy Properties:

\begin{itemize}
 \item redshift
 \item impact parameter and velocity difference to the absorber
 \item luminosity
 \item color
 \item morphology
 \item environment (in relation to other galaxies)
 \item galaxy spectral properties such as emission lines and stellar absorption 
features
\end{itemize}

Now let us try to pose a question that might allow us to choose between 
hypotheses (a) and (b) above, by using the above observables. How about: ``What 
fraction of observed Lyman $\alpha$ absorbers are physically part of the halo 
of a galaxy? (As a function of absorber column density.)''

Unfortunately, this seemingly simple query begs many questions. Is Voigt 
fitting to find absorbers in the first place meaningful (e.g. what about 
velocity cusps where there is no density enhancement)? In addition, we clearly 
need a lot more detail on what `physically part of' means. One possibility 
might be: ``bound to, and within some small physical separation from, a 
galaxy''. Even if this is sufficient, we still have to worry about galaxy 
clustering and hierarchy. Any galaxy halo will contain substructure, for 
example, the Milky Way and the Large and Small Magellanic Clouds. To which 
galaxy in that system do we give the Magellanic stream? Also, if there are High 
Velocity Clouds in the Local Group barycentre (another topic hotly debated at 
the workshop), should we give those to the Milky Way or to Andromeda?

Two final concerns undermining the validity of the question posed above are (a) 
that any galaxy sample outside of the Local Group will only contain the top end 
of the luminosity function. One can almost never be sure there isn't some 
smaller closer galaxy to any given Lyman $\alpha$ absorber. And (b), H$_0$ is 
very small (even if it is 100 km/s/Mpc), so a small uncertainty in velocity, or 
any real peculiar velocity along the Line of Sight (LOS), corresponds to huge 
distances, if mistaken for Hubble flow.

How can we modify this question to avoid some of these complications? A new 
version might be: ``Does a given model produce statistical relationships 
between `absorbers' and `galaxies' that match the observations?''. Tragically, 
this version still begs a fair number of questions. For example: can we match 
`absorbers' in the models with `absorbers' in the observations? For SPH and 
Mesh/Grid models we probably can. For the model proposing that ``All absorbers 
are part of a smooth halo around bright galaxies'', we might be able to. An 
even trickier issue is whether we can match `galaxies' in the models with real 
galaxies in the observations. For SPH and Mesh/Grid models, the answer is a 
definite `maybe'. At the moment one needs either semi-analytic add-ons, or some 
column density cut which may well not correspond to galaxies. For the ``all 
absorbers ...'' model at least, this matching is straightforward.

As a final piece of philosophy to complete this navel-gazing exercise, what do 
we do if both of the models match the currently observed statistical 
relationships (such as that between log EW and log Impact Parameter)? Clearly, 
we then need to go to higher order statistics, such as residual log (EW) vs. 
log (galaxy luminosity) or the evolution of any log EW vs. log Impact Parameter 
relationship with redshift. This approach is being followed by several groups, 
including ourselves. The other answer is the perpetual cry from observational 
astronomers: ``We need more data''. This leads into the next section describing 
a program in progress to obtain more data.

\section{New Analysis status report}

An observing program at the Canada France Hawaii Telescope (CFHT) and the WIYN 
telescope was started back in 1995. Currently the data in hand (i.e. reduced 
but not fully analyzed) are:

\noindent
WIYN Hydra Spectroscopy
 
\begin{itemize}
  \item 50-150 galaxies per field, Generally z$<$0.4
  \item 23 QSO LOS
\end{itemize}

\noindent
CFHT MOS Spectroscopy

\begin{itemize}
  \item 40-50 galaxies per mask
  \item 26 masks in 15 QSO LOS
\end{itemize}

\noindent
The basic approach for the CFHT part of the project is itemized below:

\begin{enumerate}
 \item Choose QSOs with UV spectroscopy from the HST QSO absorption line Key 
Project, giving a Lyman absorber line list.
 \item At the CFHT, image the field and do aperture photometry on the galaxies 
to generate galaxy list over a ~10 arcmin diameter region.
 \item Design  and cut masks for spectroscopy, ranking the choice of galaxies 
by their magnitudes. As usual with the CFHT MOS, this was done in real time at 
the telescope during the same run that the images were taken.
 \item Take low resolution spectra of $\sim$50 galaxies at a time to get their 
redshifts, accurate to $<$100 km/s.
 \item Design and perform statistical tests to check the models
\end{enumerate}

In order to illustrate some of the problems discussed in the section 2, we show 
below the pie diagram for the line of sight towards the quasar pair Q0107-025A 
and B. The two quasars lie off the RHS of the figure at z$\sim$1. The dotted 
lines indicate the LOS to the two quasars. The filled triangles indicate 
absorbers with EW$>$0.5A, open triangles, absorbers weaker than that. Filled 
circles are galaxies within 500 kpc (h=0.5, $\Omega$=1) of the QSO LOS, open 
circles are galaxies further from the LOS. 

\begin{figure}[ht]
\plotfiddle{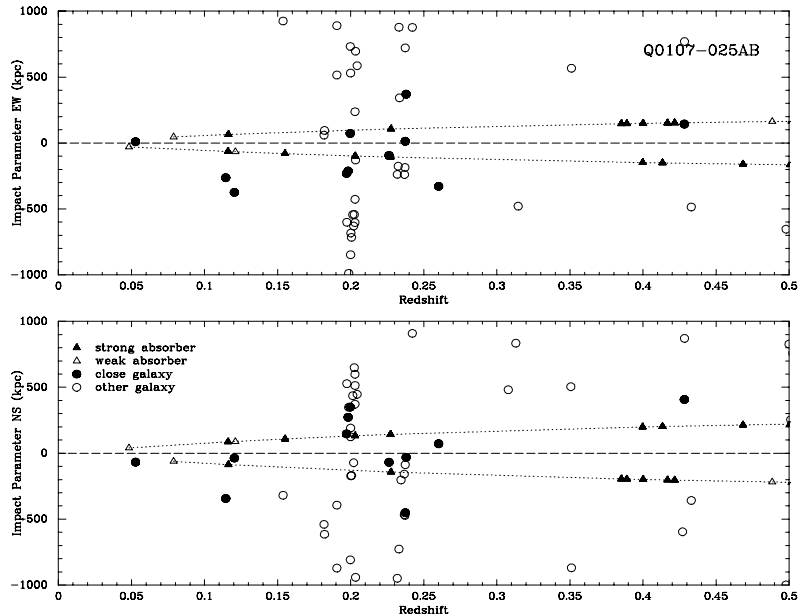}{9cm}{90}{50}{50}{198}{-18}
\caption{Pie-diagrams for the galaxies and absorbers in the LOS to the
QSO pair Q0107-025A,B. \label{smorris_1.eps}}
\end{figure}

A number of interesting associations (and lack of associations) can be seen in 
the figure:
\begin{itemize}
\item Starting simply, there is a pleasing match between a bright galaxy at 
z=0.2262, projected between the two QSOs on the sky (separation about 200 kpc 
from each LOS), and absorption seen in both LOS. Velocity differences of 480 
and 240 km s$^{-1}$ are measured between the galaxy and the absorbers. One of 
the absorbers has a high enough column density that CIV is also detected.
\item Possibly more surprisingly, there is a good match between a galaxy at 
z=0.1145 and absorption in both QSO LOS. The galaxy is 450 kpc from both LOS, 
and has a velocity difference of 400 and 480 km s$^{-1}$ from the two absorbers.
\item A complex situation is seen at z=0.2, where there is a plethora of 
candidate galaxies for association with an absorber seen in one QSO LOS but not 
the other. Three galaxies have measured redshifts within 25 km s$^{-1}$ of the 
absorber, while several others are within 500 km s$^{-1}$. There is also what 
seems to be an interacting pair of galaxies at a distance of 720 kpc and a 
velocity difference of 670 km s$^{-1}$ from the same absorber.These might be 
considered good candidates if we think tidal tails are a potential source of 
Lyman $\alpha$ absorption.
\item At a redshift of 0.2366, there are 6 galaxies all within a velocity range 
of 300 km s$^{-1}$ of each other and within a Mpc of the QSO LOS, which show no 
absorption in either LOS. The picture emerging does seem to be more like 
`weather in space' rather than well organized spherical halos.
\item Finally, we note that we have not yet identified the galaxy causing the 
Lyman limit absorption in the LOS to Q0107-025B (with corresponding absorption 
also in the LOS to Q0107-025A). Even our rather crude imaging taken through the 
CFHT MOS would have spotted a luminous (L*) galaxy on top of the QSO, so it 
seems likely that any galaxy (should it exist) which one might want to 
associate with this absorber will turn out to be a dwarf.
\end{itemize}

The purpose of the above (rather subjective) analysis of two close together 
pencil beams through space is to illustrate the difficulty in being confident 
of any assignment of a unique galaxy to a unique absorber. While single LOS may 
allow one to hope for a tidy universe with large coherent spheres of gas 
surrounding luminous and easily surveyed galaxies, there are already many 
indications that the true situation is much more complex, and therefore also 
more interesting.

\end{document}